\documentclass[iop]{emulateapj}
\usepackage{amsmath,amssymb,graphicx,color,longtable,natbib}

\slugcomment{}
\shorttitle{2.5 billion microlensing light curves}
\shortauthors{Vernardos et al.}

\begin{document}
\normalsize

\title{GERLUMPH Data Release 2: 2.5 billion simulated microlensing light curves}

\author{G.~Vernardos\altaffilmark{1}, C.J.~Fluke\altaffilmark{1}, N.F. Bate\altaffilmark{2,3}, D. Croton\altaffilmark{1}, and D. Vohl\altaffilmark{1}}
\affil{$^1$Centre for Astrophysics \& Supercomputing, Swinburne University of Technology, PO Box 218, Hawthorn, Victoria, 3122, Australia}
\affil{$^2$Sydney Institute for Astronomy, School of Physics, A28, University of Sydney, NSW, 2006, Australia}
\affil{$^3$School of Physics, The University of Melbourne, Parkville, VIC 3010, Australia}

\begin{abstract}
In the upcoming synoptic all--sky survey era of astronomy, thousands of new multiply imaged quasars are expected to be discovered and monitored regularly.
Light curves from the images of gravitationally lensed quasars are further affected by superimposed variability due to microlensing.
In order to disentangle the microlensing from the intrinsic variability of the light curves, the time delays between the multiple images have to be accurately measured.
The resulting microlensing light curves can then be analyzed to reveal information about the background source, such as the size of the quasar accretion disc.
In this paper we present the most extensive and coherent collection of simulated microlensing light curves; we have generated $>2.5$ billion light curves using the GERLUMPH high resolution microlensing magnification maps.
Our simulations can be used to: train algorithms to measure lensed quasar time delays, plan future monitoring campaigns, and study light curve properties throughout parameter space.
Our data are openly available to the community and are complemented by online eResearch tools, located at \url{http://gerlumph.swin.edu.au}.
\end{abstract}


\section{Introduction}
\label{sec:intro}
Gravitationally lensed quasars are unique natural laboratories for exploring the physics of extreme environments in the Universe.
Microlensed quasars in particular represent the only systems known so far where properties such as the temperature profile of the quasar accretion disc can be probed.

Quasar microlensing was first observed as uncorrelated light curve variability attributed to compact stellar mass objects within the galaxy--lens that lie close to the line of sight \citep{Irwin1989,Chang1979,Chang1984,Paczynski1986,Kayser1986,Schneider1987a}.
From this it has been possible to derive constraints on the accretion disc size and temperature profile in a number of systems.

Accretion disc size estimates for wavelengths between 0.07 and 2 microns vary in the range $14.5 \lesssim {\rm log} \left( r_{1/2} \right) \lesssim 17.3$, where $r_{1/2}$ is the half light radius measured in cm \cite[e.g. see][and references therein for results on 25 systems]{Pooley2007,Morgan2010,Blackburne2011,Munoz2011,Jimenez2014}.
Temperature profiles are also found to be rather flat, i.e. independent of wavelength \citep[e.g. see][]{Blackburne2011,Jimenez2014}.
These results are in disagreement with the thin disc theory \citep{Shakura1973}, which predicts sizes down to an order of magnitude smaller and temperature profiles with a $4/3$ power law dependence on wavelength.

Another experiment that can be performed using gravitationally lensed quasars is measuring the value of Hubble's constant, $H_0$ \citep{Refsdal1964b}.
This has been performed in at least 24 multiply imaged systems \citep[see][and references therein]{Tortora2004,Paraficz2010,Eulaers2011,RathnaKumar2014} to yield values in the range $50-100$ km s$^{-1}$ Mpc$^{-1}$.
The wide range of values, which is less accurate than results obtained from other methods \citep[e.g. Cepheid variability;][]{Riess2011,Freedman2012}, is due to uncertainties and degeneracies of the lens potential \citep{Kochanek2002,Kochanek2004b,Oguri2007,Schneider2013,Suyu2014}.

In order to obtain light curves suitable for $H_0$ and accretion disc studies, lens systems have to be monitored for periods of months, or even years \citep[e.g. see][]{Gott1981,Kundic1997,Eigenbrod2005,Fohlmeister2008}. Moreover, accurate observational constraints, such as the positions and photometry of the multiple images, are required to produce accurate models of the galaxy--lens.
Extracting the correct time delays can be further complicated by the onset of microlensing, which introduces additional uncorrelated variability between the multiple images \citep[e.g. see][]{Hojjati2013,Tewes2013}.
In fact, microlensing is expected to be taking place in all gravitationally lensed quasars, although with varying strength.

The effect of microlensing is usually modeled using a magnification map: a pixellated map of the magnification in the source plane induced by the foreground microlenses (see Figure \ref{fig:locations}).
Such maps are produced using different implementations of the inverse ray--shooting technique \citep[e.g.][]{Kayser1986,Wambsganss1999,Kochanek2004,Thompson2010,Mediavilla2011b,Vernardos2014b}.

Due to the relative motion of observer, lens, and source, the source appears to move across the map crossing regions of high and low magnification and producing a variable light curve.
Simulated light curves have been used to study high magnification events \citep{Witt1990,Yonehara2001,Shalyapin2002}, autocorrelation functions \citep{Seitz1994b,Lewis1996,Wyithe2002a}, and derivative distributions \citep{Wyithe1999}.
\cite{Kochanek2004} developed a quantitative Bayesian analysis technique which has been used to study the background quasar \citep[e.g.][]{Morgan2010,Dai2010} and the foreground galaxy--lens \citep[e.g.][]{Mosquera2013}.

From the $\sim90$ known multiply imaged quasars \citep[][]{Mosquera2011b}, $\sim25$ have been studied using microlensing techniques, either as single objects or in groups of a few.
However, this is about to change due to the upcoming synoptic all--sky survey facilities, like the Large Synoptic Survey Telescope \citep[LSST;][]{LSST2009}, which are expected to discover thousands of new lensed quasars suitable for microlensing studies \citep{Oguri2010}.
Moreover, the high cadence of observations from these instruments will provide nearly effortless monitoring.
These data are expected to increase the accuracy of $H_0$ measurements \citep[e.g.][]{Coe2009,Dobke2009}, and improve our current techniques for constraining quasar structure \citep[e.g.][]{Sluse2014}.
It is crucial for microlensing to move from single object to parameter space studies.

As a theoretical counterpart of future quasar microlensing observational campaigns, the Graphics Processing Unit--Enabled High Resolution cosmological MicroLensing parameter survey\footnote{\url{http://gerlumph.swin.edu.au}} \citep[GERLUMPH;][]{Vernardos2014b} has already generated $>70,000$ microlensing magnification maps, the largest and most complete collection yet produced.
The parameter space of convergence $\kappa$, shear $\gamma$, and smooth matter fraction $s$, (see next section) is covered in unprecedented detail, allowing for comprehensive explorations of microlensing properties \cite[e.g.][]{Vernardos2014c}.

In this paper we present how we used the GERLUMPH maps to generate $>2.5$ billion simulated microlensing light curves, the largest and most extensive set of light curves available to date.
Our approach spreads the modelling effort all over the parameter space rather than focusing it on a single object, e.g. the method of \cite{Kochanek2004} may extract up to $10^6$ light curves from a single source profile and magnification map, while we are extracting 2000 light curves per source profile from 51,127 maps covering the parameter space (see next section).
Hence, our data is not designed for detailed modelling of individual objects but it is designed for studying the robustness of, and degeneracies in, models for many individual objects across the parameter space.
Using these simulations to unveil systematic errors introduced in the modelling process will hopefully lead to better measurements of $H_0$ and accretion disc constraints.

We present our approach to extracting simulated microlensing light curves in Section \ref{sec:method}.
Our data are described in Section \ref{sec:results} and are openly accessible for download, complemented by online analysis tools that we introduce in Section \ref{sec:tools}.
Finally, we conclude our paper and present future prospects in Section \ref{sec:discussion}.

\section{Approach}
\label{sec:method}
We have used 51,227 microlensing magnification maps from the GERLUMPH online resource, generated using the {\tt GPU-D} \citep{Vernardos2014b,Thompson2014} direct Graphics Processing Unit (GPU) implementation of the inverse ray--shooting technique \citep{Kayser1986}.
The map resolution is 10,000 pixels on a side, the map width is set to 25 Einstein radii ($R_{\rm Ein}$, see below), and the mass of the microlenses are fixed at 1 M$_{\odot}$.
The maps are extracted from the region of parameter space with $0<\kappa<1$ and $0\leq\gamma<1.3$, which contains most of the models of the galaxy--lens of the currently known systems \citep[see figure 2 of][or see the relevant GERLUMPH online tool\footnote{\url{http://gerlumph.swin.edu.au/macromodels/}}]{Vernardos2014a}.

The smooth matter fraction is defined as $s=\kappa_{\rm s}/\kappa$, where $\kappa_{\rm s}$ is the contribution to the total convergence by the smooth matter component.
For each $\kappa,\gamma$ combination we use 10 maps with different smooth matter content: $0 \leq s \leq 0.9$ in steps of $0.1$.
The collection of maps used in this study is contained within the slightly larger set of maps used previously in \cite{Vernardos2014c}.
Properties of the GERLUMPH high resolution maps are examined in \cite{Vernardos2014a}, while a description of the GERLUMPH data, infrastructure, and tools, can be found in \cite{Vernardos2014b}.

The characteristic microlensing scale length in the source plane is the Einstein radius:
\begin{equation}
\label{eq:rein}
R_{\rm Ein} = \sqrt{ \frac{D_{\rm os}D_{\rm ls}}{D_{\rm ol}} \frac{4G\langle M \rangle}{c^2} } \, ,
\end{equation}
where $D_{\rm ol}$, $D_{\rm os}$, and $D_{\rm ls}$, are the angular diameter distances from observer to lens, observer to source, and lens to source respectively, $\langle M \rangle$ is the mean mass of point--mass microlenses, $G$ is the gravitational constant, and $c$ is the speed of light.
A typical range of values for $R_{\rm Ein}$ can be obtained from the sample of 87 lensed quasars compiled by \cite{Mosquera2011b}: $5.11 \pm 1.88 \times 10 ^{16}$cm.
Although these authors have used estimates of the lens redshift in a number of cases, their result is consistent with the CASTLES\footnote{\url{http://www.cfa.harvard.edu/castles/}} sample of 59 systems with both lens and source redshifts measured \citep[5.35$\pm$2.20$\times$10$^{16}$cm,][]{Falco2001}.
In the following, we use the mean from the \cite{Mosquera2011b} sample as a typical value for $R_{\rm Ein}$.
This leads to a pixel size of the high--resolution GERLUMPH maps of $\sim1.28 \times 10^{14}$cm.

Because the microlensing effect depends weakly on the shape of the underlying accretion disc brightness profile \citep{Mortonson2005}, the half light radius, $r_{1/2}$, is a convenient way to parametrize a range of disc models consistently.
We model the quasar source profile as a face--on\footnote{Quasars are thought to be viewed mostly face--on within the current framework of the unified model for Active Galactic Nuclei \citep{Antonucci1993,Urry1995}. This is supported by microlensing results for at least one system \citep[Q2237$+$0305;][]{Poindexter2010b}. However, depending on the (unknown) width of the dusty torus surrounding the central regions of the quasar, the viewing angle could be very different to face on. Since microlensing behaviour depends only weakly on shape, a disc that is not viewed face--on will just appear smaller.} Gaussian disc, i.e. $I\left( r \right) = \mathrm{exp}\left(-r^2/2\sigma^2\right)$, of varying size $\sigma$ that is related to $r_{1/2}$ through
\begin{equation}
\label{eq:sigma_rhalf}
r_{1/2} = 1.18 \sigma.
\end{equation}
We truncate the profile at $r=3\sigma$, which is the radius containing $99.7$\% of the total brightness.
Therefore, the total size, or diameter, of the profile, $d$, which is used to determine the size of the effective map (see below), is equal to $6\sigma$.
We consider 25 profiles of varying size $d$: from $2\times10^{15}$cm to $2\times10^{16}$cm ($\sim0.8$ to $\sim8$ light days) in steps of $2\times10^{15}$cm, and from $2\times10^{16}$cm to $1.7\times10^{17}$cm ($\sim8$ to $\sim65$ light days) in steps of $1\times10^{16}$cm.
Consequently, we cover the range $14.6 \lesssim {\rm log} \left( r_{1/2} \right) \lesssim 16.5$ which contains most of the current microlensing size estimates of the accretion disc (see Section \ref{sec:intro}).

As it will be shown in the next section, including wider profiles produces large amounts of additional data without significant change in the light curves.
Such a behaviour is expected because the microlensing effect is more prominent for smaller sources \citep[e.g.][]{Wambsganss1991}, while the magnification of large sources tends to the macro--magnification (see Equation \ref{eq:muth}).

Light curves extracted from the original GERLUMPH maps hold magnification information for ``pixel--size'' sources, i.e. sources that are smaller than the pixel size of the maps.
To get information for a source profile with finite size one has to convolve it with the maps.
This can be achieved using the convolution theorem and a GPU to accelerate computing Fourier transforms \citep[see][for a detailed description of this technique]{Vernardos2014b}.
Assuming the map is periodic, which is required by the Fourier transform, leads to spurious magnification values around the edges of the convolved map.
The size of these regions is equal to half the source profile size, i.e. $d/2$.
The largest profile we use is 1332 pixels wide, therefore, by disregarding areas of 700 pixels around the edges of the map we are not affected by spurious magnification values.
The resulting effective convolved maps, from which the light curves are extracted, have a width of 8600 pixels, or $21.5 R_{\rm Ein}$.

\begin{figure}
\includegraphics[width=0.47\textwidth]{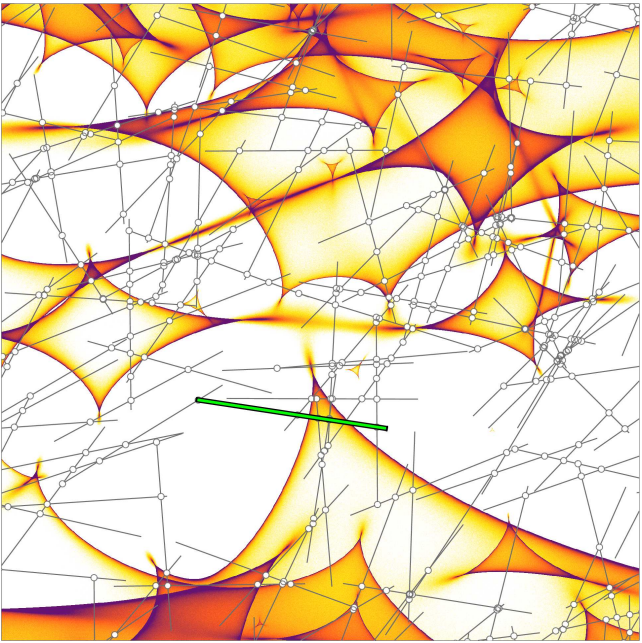}
\caption{Central 2000--pixel (5 $R_{\rm Ein}$) wide region of a magnification map with $(\kappa,\gamma,s) = (0.52,0.36,0.5)$, produced by 5797 microlenses. The locations of the sampled light curves and their intersection points are shown in grey. The magnification along the thick (green online) curve is shown in Figure \ref{fig:data} for 25 different source profiles.\label{fig:locations}}
\end{figure}

We extract 2,000 light curves from each effective map.
Ideally, this number would be as high as possible, however, increasing it further will produce much more data than what we can currently store.
It turns out that this number is sufficient to adequately sample the underlying magnification probability distribution of the magnification maps (see next Section).

The light curve locations are randomly selected once, and then used for all the maps.
The microlens positions for generating the GERLUMPH maps are set randomly using different random seeds for each map.
Therefore, the locations of caustic networks between maps are not correlated and the fixed light curve locations do not bias the actual light curves extracted.
This approach reduces the number of parameters that we have to keep track of, and avoids issues with generating random numbers, e.g. in case of reproducing our results using different compiler versions.
In Figure \ref{fig:locations}, we show a 2000--pixel--wide region from the center of a fiducial magnification map with the locations of the extracted light curves.

The light curve length is set to $1.5 R_{\rm Ein}$ and sampling length to $0.0025 R_{\rm Ein}$ (this is set by the resolution of our maps); using the values for $R_{\rm Ein}$ and the effective source velocity from \cite{Kochanek2004} for Q2237$+$0305 the light curve would be $\sim22$ years long, sampled every $\sim13$ days.
Ideally, longer light curves would be desired, however, further increasing the light curve length will produce much more data than what we can currently store.
A length of  $1.5 R_{\rm Ein}$ is adequate to capture the microlensing variations ($R_{\rm Ein}$ is the typical scale length for the onset of microlensing) and corresponds to $\sim3\sigma$ of the Gaussian distribution we used to model our largest source profile.

The actual number of pixels crossed varies according to the orientation of the light curve: 600 pixels are crossed in the horizontal or vertical direction, while only 425 pixels are crossed at an angle of $45^{\circ}$.
Obviously, in the latter case some pixels are sampled more than once leading to the appearance of short flat parts, or steps, in the light curve.
This can also arise due to the orientation of the light curve with respect to a caustic and is a pathological behaviour due to the finite resolution of a pixellated magnification map.
Some interpolation prescription between neighbouring pixels would resolve this and make the light curves smoother.
However, we currently avoid making any assumptions on such smoothing procedures and provide the raw sampled data.

Our approach to pixel sampling lies between Bresenhams' algorithm \citep{Bresenham1965} and a supercover algorithm \citep[e.g.][]{Andres1997}, which both belong to the digital differential analyzer (DDA) class of algorithms used in computer graphics for rasterization of geometrical shapes.
In fact, infinitely reducing the scaling length would lead exactly to a supercover pixel sampling, i.e. sampling of all the pixels a line crosses.
The maximum amount of light curve information extracted from a magnification map would be possible by calculating each portion of a curve that lies in a given pixel.
However, our choice of sampling avoids the need for complex data structures, increased data size (two values required for each pixel), and resembles more the observational data which inevitably produce discrete samples.

\subsection{The data}
We have carried out $>1.28$ million convolutions across 5 days on a GPU supercomputer.
Each combination of map, profile, and $R_{\rm Ein}$ is stored in an indexed database; the index of each entry points to a directory in a flat file system that holds all output \citep[more details on the choices we made for the backend database and file storage can be found in][]{Vernardos2014b}.
Each light curve consists of 600 sampled magnification values which are represented as 32--bit floats.
The 2000 light curves per convolution are stored in a binary file of $\sim4.6$ MB in size.
We have generated a total of $\sim2.5\times10^9$ light curves, corresponding to $5.6$ TB of data.
All our data are freely accessible to analyze and download using the eTools described in Section \ref{sec:tools}.

\section{Results}
\label{sec:results}
As a first test, we consider maps for pixel--size sources and extract the magnification probability distribution (MPD) based purely on the light curve magnification values from 2000 light curves.
Then, we compare this light curve MPD to the full MPD extracted from the same map region using the Kolmogorov--Smirnov test with the null hypothesis that the two distributions are the same.
We find that $\sim8$\% of the 51,227 tests fail with a p--value $<0.05$.
We note that the light curves intersect at 5,995 pixel locations (same for every convolved map), which are counted more than once in the calculation of the MPD, but given the low number of failed tests we choose to ignore this.

More than 95\% of the failed tests occur for maps produced by $< 10^4$ microlenses.
This means that the caustic networks are less dense and a higher number of light curves would be needed to better probe the underlying map MPD.
Indeed, increasing the number of light curves to 4000 and 8000 leads to $\sim6.5$\% and $\sim4$\% of failed tests respectively (but increases the size of the data products by factors of 2 and 4).
Finally, it is expected that when a pixel--size source map is convolved with a source profile, the caustic networks, and the corresponding MPD, are smoothed out.
We perform the same MPD comparisons using 2000 light curves from convolved maps, and find that even the smallest source profile, i.e. $d=2\times10^{15}$cm or $\sim 0.04 R_{\rm Ein}$, leads to $<1$\% of failed tests.

The Kolmogorov--Smirnov test is only one method of assessing the usefulness of any sub--sample from a population.
The test is more sensitive to differences near the peak of the distribution, rather than at the extremes where the individual probabilities are low \citep[see section 4.1 of][]{Vernardos2014c}.
We reiterate that it is straightforward to generate additional lightcurves per map, and indeed more independent magnification maps, but it is non-trivial to store them.

\begin{figure}
\includegraphics[width=0.47\textwidth]{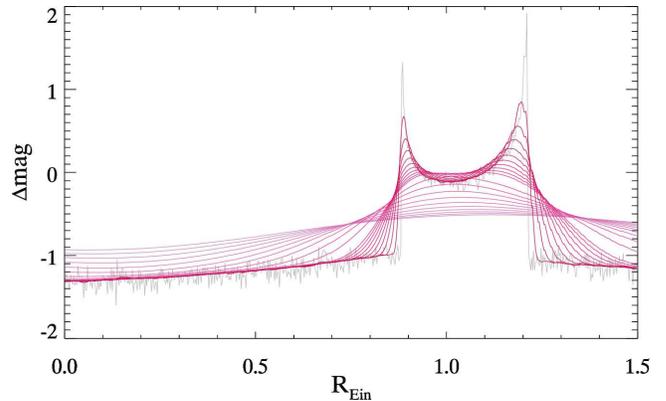}
\caption{Magnification values, expressed in units of $\Delta \mathrm{mag}$ (Equation \ref{eq:dmag}), along the trajectory shown in Figure \ref{fig:locations}. The thick black line (magenta online) corresponds to the smallest source size for which the microlensing induced fluctuations are the most prominent. The double peaks disappear for a profile with a size of $\sim 0.4 R_{\rm Ein}$, roughly equal to the double peak separation. The pixel--size source light curve is shown in grey.\label{fig:data}}
\end{figure}

In Figure \ref{fig:data} we plot the magnification values along the trajectory shown in Figure \ref{fig:locations} for 25 different profile sizes.
We convert the magnification of the unlensed source luminosity into magnitude change with respect to the macro--magnification produced by the galaxy--lens without the additional effect of microlensing:
\begin{eqnarray}
\label{eq:dmag}
\Delta \mathrm{mag} &=& 2.5 \, \mathrm{log} \left( \frac{L^{total}}{L} \right) - 2.5 \, \mathrm{log} \left( \frac{L^{macro}}{L} \right) \nonumber \\
                    &=& 2.5 \, \mathrm{log} \left( \mu \right) - 2.5 \, \mathrm{log} \left( \mu_{\rm th} \right) \nonumber \\
                    &=& 2.5 \, \mathrm{log} \left( \frac{\mu}{\mu_{\rm th}} \right),
\end{eqnarray}
where $L$ is the unlensed source luminosity, $L^{total}$ is the total magnified flux, $L^{macro}$ is the magnified flux due to the galaxy--lens only, and the theoretical macro--magnification is:
\begin{equation}
\label{eq:muth}
\mu_{\rm th} = \frac{1}{(1-\kappa)^2-\gamma^2}.
\end{equation}
For small source sizes, the light curve has a characteristic double--peaked shape, with the peaks corresponding to the caustic crossing events shown in Figure \ref{fig:locations}.
As the profile size gets larger the two peaks are smoothed out and disappear for the profile with size $2\times10^{16}$cm, or $\sim 0.4 R_{\rm Ein}$, which corresponds roughly to the length of the light curve that lies within the caustic.
For even larger profiles the light curve becomes almost flat and microlensing fluctuations are much less prominent as expected.

In the original implementation of {\tt GPU-D} \citep{Thompson2010}, we used a random distribution of light rays.
Compared to the alternative, i.e. a regular grid, this was a compromise that helped amortize the computational cost of additional GPU-computation while allowing us to explore billion
lens configurations with abitrary numbers of source-plane pixels.
To this end, light ray positions were generated on the CPU (host memory) while the light ray deflections were being calculated on the GPU.
As a result, there is an uncertaininty in the magnification values from each map pixel: counting $N_{i,j}$ rays in a given pixel is accompanied by an approximately $\sqrt{N_{i,j}}$ error, i.e. Poisson-like.
The code comparisons made by \cite{Bate2010} showed that the these per-pixel magnification errors were small, and become less significant as the maps are convolved with realistic source profiles.

Low magnifications correspond to low ray counts and consequently larger errors with respect to the value itself (i.e. $N^{-1/2}$).
Therefore, we expect larger fluctuations in the low magnification parts of a simulated light curve for sources smaller than the size of the map pixels.
Indeed, such a behaviour is observed for the pixel--size source light curve shown in Figure \ref{fig:data}.
As soon as the pixel--size source map is convolved with a profile - even the smallest profile - the fluctuations disappear and the light curve becomes smoother.

\begin{figure}
\includegraphics[width=0.47\textwidth]{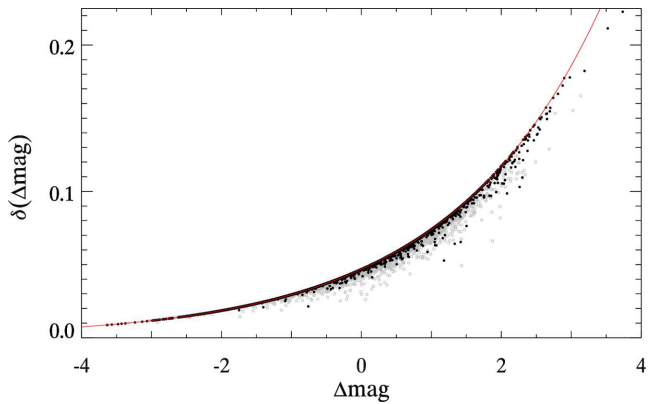}
\caption{Magnification errors in units of $\Delta\mathrm{mag}$ (same as y--axis of Figure \ref{fig:data}). The exact error for convolved magnification values (Equation \ref{eq:exact_error}) is computed in 100 locations within a magnification map; 130 maps are used spanning the parameter space examined here. The results for a convolution kernel 16 pixels wide are shown as filled black circles, while open grey circles correspond to a kernel with 158 pixels on a side. Different errors for the same magnification (vertical spread of the points) are produced due to different neighbouring pixels within the range of the convolution kernel, e.g. near a caustic. The solid line (red online) is the approximation of Equation (\ref{eq:assumed_error}); values of $\langle \mu \rangle$ and $\langle N \rangle$ from any of the maps produce practically the same curve.\label{fig:errors}}
\end{figure}

The error in the magnification $\mu_{i,j}$ of a given map pixel is
\begin{equation}
\label{eq:error}
\delta\mu_{i,j} = \sqrt{N_{i,j}} \frac{\langle \mu \rangle}{\langle N \rangle},
\end{equation}
where $\langle N \rangle$ is the average number of rays per map pixel, and $\langle \mu \rangle$ is the average magnification per pixel, which can be computed by the total number of rays in a map and the number of rays that would have reached the map if there was no lensing \citep[see the Appendix of][for more details]{Vernardos2013}.
For convolved maps, this error propagates through the convolution formula:
\begin{eqnarray}
\label{eq:exact_error}
\delta\mu_{i,j}' &=& \sum_{m,n=-s}^{s}  k_{m,n} \delta\mu_{i+m,j+n}  \nonumber \\
				 &=& \frac{\langle \mu \rangle}{\langle N \rangle} \sum_{m,n=-s}^{s}  k_{m,n} \sqrt{N_{i+m,j+n}}
\end{eqnarray}
where we have assumed a square, normalized kernel $k$, with a size of $2s$ pixels, centered over the $i,j$-th map pixel.
Keeping track of the exact error of the convolved magnification values means convolving the ``error'' map with the source profile and storing an additional value for each light curve pixel.
Such an approach would double the number of convolutions performed and data that have to be stored.
To avoid this, we approximate the error given by Equation (\ref{eq:exact_error}) by assuming that the Poisson statistics hold for the convolved ray--count map.
In this case, the error for the convolved magnification is:
\begin{equation}
\label{eq:assumed_error}
\delta\mu_{i,j}' = \sqrt{N_{i,j}'} \frac{\langle \mu \rangle}{\langle N \rangle},
\end{equation}
where $N_{i,j}'$ is not necessarily an integer ray--count anymore.
Taking into account that $\mu_{i,j}' = N_{i,j}' \langle \mu \rangle / \langle N \rangle$, we end up with the final expression for the convolved light curve error:
\begin{equation}
\label{eq:final_error}
\delta\mu_{i,j}' = \sqrt{\mu_{i,j}'\frac{\langle \mu \rangle}{\langle N \rangle}}.
\end{equation}
In other words, we approximate the error propagated by the convolution with an assumed Poisson error of the convolved pixel in question.

In Figure \ref{fig:errors}, we compare the above approximation to the actual error, calculated in randomly sampled pixels from a representative set of maps.
We see that our approximation is in fact a maximum error of the magnification, which originates from the observation that
\begin{equation}
\label{eq:inequality}
\sqrt{ \sum_{m,n=-s}^{s}  k_{m,n} N_{i+m,j+n} } \geq \sum_{m,n=-s}^{s}  k_{m,n} \sqrt{N_{i+m,j+n}}.
\end{equation}
This inequality depends on the actual convolution kernel used, however, and proving it is beyond the purpose of this paper.
It suffices to say here that the Gaussian profiles we are using are peaked at the location of the $i,j$-th pixel; if profiles that are peaked away from the central pixel are used, then our approximation is most likely going to fail.

Thus far we have assumed a fixed physical size for the profiles, and the convolutions have been performed for a fixed value of $R_{\rm Ein}$ (see Section \ref{sec:method}).
However, what if a different value of $R_{\rm Ein}$ is used, e.g. to study specific lensed systems?
In this case, the GERLUMPH light curve data can still be used, but caution has to be taken on how to scale the source sizes correctly.
For example, if we apply the GERLUMPH maps to quasar Q2237$+$0305 that has $R_{\rm Ein} = 1.81\times 10^{17}$cm for 1 M$_{\odot}$ microlenses \citep{Mosquera2011b}, our largest source profile, $1.7 \times 10^{17}$cm, would correspond to a 376--pixel Gaussian kernel.
The closest convolution kernel to this size is 392 pixels, which corresponds to a $5\times10^{16}$cm profile and $R_{\rm Ein} = 5.11 \times 10^{16}$cm.
Of course, the closer the size of the source profile in pixels is to one of the convolution kernels the more similar the light curves will be.
The case with equal kernel sizes would lead to identical light curves, meaning that the scaled light curves would be exact, although a different value of $R_{\rm Ein}$ was used.

Variations of the value of $R_{\rm Ein}$ due to other reasons, e.g. uncertainties in the redshifts of lens and source or different mass of microlenses, lead to similar scaling of the source profile sizes.
We demonstrate this with an example of a variation in the Hubble constant.
The value of $R_{\rm Ein}$ used in this paper, i.e. $5.11 \pm 1.88 \times 10 ^{16}$cm \citep{Mosquera2011b}, was obtained assuming a Universe with $\Omega_m=0.3$, $\Omega_{\Lambda}=0.7$, $\Omega_{k}=0$, and $H_0=72$ km s$^{-1}$ Mpc$^{-1}$.
The angular diameter distances appearing in Equation (\ref{eq:rein}), $D_{A}$, depend linearly on the line-of-sight comoving distance\footnote{This equation is for an object at redshift $z$ and $\Omega_{k}=0$. Getting the angular diameter distance $D_{A;1,2}$ between two objects at redshifts $z_1$ and $z_2$ depends on the geometry of the Universe and in general $D_{A;1,2} \neq D_{A;2} - D_{A;1}$. However, for $\Omega_{k}=0$, $D_{A;1,2} = \frac{D_{C;2}-D_{C;1}}{1+z_2}$.}, $D_{C}$, as
\begin{equation}
D_{A} = \frac{D_{C}}{1+z},
\end{equation}
which in turn is proportional to the Hubble distance, $D_{H} = c/H_0$.
This leads to
\begin{equation}
\frac{R_{\rm Ein}}{\delta R_{\rm Ein}} = \frac{-2 H_0}{\delta H_0},
\end{equation}
that relates the error in $H_0$ with the error in the derived $R_{\rm Ein}$ value.
Our profile sizes can then be scaled according to the discussion in the previous paragraph.
Similar reasoning could lead to a different relation for the dependence of $R_{\rm Ein}$ on $H_0$ for different cosmological parameters.

\section{eTools}
\label{sec:tools}

\subsection{Accessing the data}
The GERLUMPH light curve data can be openly accessed online.
To this purpose, we have extended the {\tt getquery} eTool described in detail in \citet[][Section 4 and Figure 7]{Vernardos2014b}.
Detailed instructions and help tips on how to download the data are provided on the GERLUMPH website\footnote{visit \url{http://gerlumph.swin.edu.au/guide/} and \url{http://gerlumph.swin.edu.au/getquery/}.} throughout the various stages of the process.

The 2000 light curves for each map are stored in a binary file (\texttt{lc\_data.bin}), which is compressed using a standard Unix tool (e.g. \texttt{gzip} or \texttt{bzip2}).
The downloaded files are grouped in indexed directories for each map and source profile.
The values of $\langle \mu \rangle$ and $\langle N \rangle$, required to calculate the errors in the magnification values, are stored in a metadata file (\texttt{mapmeta.dat}) in the map directories.
Finally, at the root directory there are two reference files for looking up the map and profile indices (\texttt{mINDEX.txt} and \texttt{pINDEX.txt}) and their properties, e.g. $\kappa$, $\gamma$, profile size, etc.

\begin{figure*}
\includegraphics[width=\textwidth]{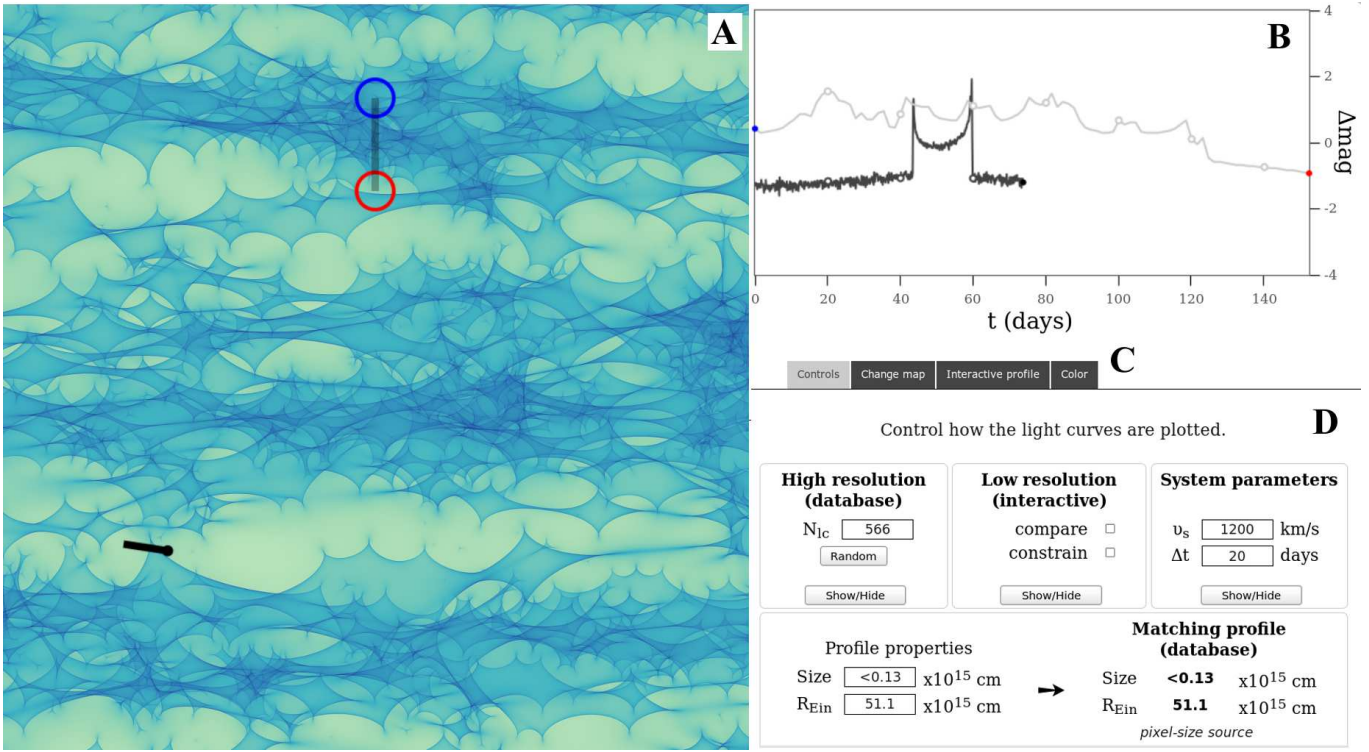}
\caption{Screenshot of the \texttt{GIMLET} tool, located at \protect\url{http://gerlumph.swin.edu.au/tools/gimlet}. Detailed instructions and help in using the GIMLET tool can be found online. Here we outline its main features: (A) Scaled down version of a GERLUMPH map, with the locations of the high and low resolution light curves. The low resolution light curve can be moved and rotated interactively by the user. (B) The corresponding high (black) and low (grey) resolution light curve variations, plotted as a function of length, measured in $R_{\rm Ein}$, or time measured in days. (C) Panels that allow the map, source profile, and color coding of the map to be changed (see Section \ref{sec:gimlet} for more details). (D) The control panel for displaying the light curves and enabling sampling. \label{fig:gimlet}}
\end{figure*}

\subsection{The \texttt{GIMLET} tool}
\label{sec:gimlet}
Simulated light curves from the GERLUMPH maps can be inspected using the GERLUMPH Interactive Microlensing Lightcurve Extraction Tool (\texttt{GIMLET}):
\begin{center}
\url{http://gerlumph.swin.edu.au/tools/gimlet/}
\end{center}
which is openly available online.
This is an exploration and planning tool, whose main goals are:
\begin{itemize}
\item [1.] To plot high resolution light curves and show their location on the corresponding magnification maps.
\item [2.] To plot low resolution light curves for comparison, which can be extracted interactively in real time.
\item [3.] To demonstrate the effects of light curve sampling.
\end{itemize}
The interactive light curve is not extracted from the full resolution magnification map, as this would consume a lot of computational resources due to the size of each magnification map (381 MB).
Instead, a scaled-down version of the map is used that has been precomputed and stored in the GERLUMPH database \citep[the \texttt{sample.png} file, see][]{Vernardos2014b}.
The spatial resolution of this map icon is set to 1000$\times$1000 pixels, but it can be reduced according to the resolution of the monitor used to view the webpage.
The magnification, in units of $\Delta \mathrm{mag}$ (Equation \ref{eq:dmag}), is binned in 256 bins in the range $[-4,4]$ (values outside this range, which are extreme and rare, are set equal to the range limits).
These approximations are used to give a real-time feeling to generating light curves.
To obtain high quality data users are advised not to use the GIMLET tool and instead download the high resolution light curve data.

In Figure \ref{fig:gimlet} we show a screenshot of the \texttt{GIMLET} tool webpage.
The panel shown, ``Controls'', contains the basic features and functionality.
High and low resolution light curves can be displayed and compared.
The length of the light curves is measured in units of $R_{\rm Ein}$, but it is possible to change to units of time (days) by specifying a value for $R_{\rm Ein}$ and for the effective velocity of the source, $\upsilon_s$.
A time interval, $\Delta t$, can be set to simulate a light curve cadence of observations.
Other functions can be performed in the remaining panels: any of the GERLUMPH maps can be loaded (``Change map''), the source profile for the interactive low resolution light curves can be modified interactively (``Interactive Profile''), and the color scheme used on the displayed map can be changed (``Color'').

\subsubsection{Implementation}
The content of the tool webpage is rendered by a PHP script that handles all the communication with the database.
JavaScript functions and the jQuery\footnote{\url{http://jquery.com/}} library are used to further manipulate the elements of the webpage: reload the high resolution light curve data, extract the interactive light curve, and handle all the button functions.
The Flot\footnote{\url{http://www.flotcharts.org}} library is used to plot the light curves.
The map is rendered in a HTML5\footnote{\url{http://www.w3.org/TR/html5/}} canvas element and the KineticJS\footnote{\url{http://kineticjs.com/}} library is used to plot the light curve location on the map.
The color is applied to the map pixels using the WebGL\footnote{Web Graphics Library: \url{www.khronos.org/webgl}} JavaScript application programming interface, that allows for GPU acceleration at the user end.

\section{Summary and Discussion}
\label{sec:discussion}
We have used the GERLUMPH magnification maps to produce $>2.5$ billion simulated microlensing light curves.
Our data are publicly available for download from the GERLUMPH server.
We also release an online exploration and planning tool for plotting the high resolution light curves presented here and extracting interactive low resolution light curves in real time.
Our goal is to provide an extensive and consistent set of light curves to be used as a benchmark for future parameter space and individual system microlensing studies.
We described numerical errors of our data and scaling of the source profiles with respect to values of $R_{\rm Ein}$ and $H_0$ in Section \ref{sec:results}.

Such a complete set of light curves is of high relevance to microlensing studies of large numbers of lensed quasars in the upcoming all-sky survey era of astronomy.
While our light curve data cannot be used to fit observed light curves for single systems, they are designed to test the robustness of, and degeneracies in, such techniques for many individual objects in the parameter space.
In this way, we will be able to discover potential systematic errors introduced in the modelling process of microlensed quasars, that will hopefully lead to better measurements of $H_0$ and accretion disc constraints.

Our results can be used to train machine learning algorithms for measuring time delays \citep[e.g.][]{CuevasTello2006,Hirv2011}.
Our simulations do not try to reproduce any specific observed light curve directly -- instead, they cover a large range of possible, yet currently unobserved scenarios.
As such, they are highly suitable as an input to unsupervised machine learning algorithms as part of the process of automatically discovering and classifying thousands of new lensed quasars set to be discovered in future synoptic surveys.
For example, they are ideally suited for the time delay challenge \citep[TDC;][]{Dobler2013,Liao2014}, a collaborative community effort that uses mock observations of multiply imaged quasars and attempts to measure the time delays using various techniques.
The GERLUMPH light curves are available at high resolution and for a wide range of source and lens model parameters that can be matched to those of the TDC mock observations.

Additional machine learning and data mining approaches can be used to explore our simulated light curve dataset \cite[e.g. see][]{Ball2010,Ivezic2014}.
This could be done by calculating basic statistical properties, e.g. mean, median, variation, etc, or more advanced properties, like the power spectrum, the number of peaks in the light curves, their prominence and/or separation, etc.
The entire sample of light curves could then be classified using such a metric as a new way of exploring and understanding lens and source model specific degeneracies.
Given the large size of the data, i.e. 2.5 billion light curves corresponding to $5.6$ TB of data, the high dimensionality of the parameter space, and the many metrics and classification techniques that can be used, this task will be the topic of future work.

We have focused on the size as the most significant source factor for microlensing \citep[][]{Mortonson2005}.
However, theoretical studies could be envisaged using more complicated source profiles (inclined discs, biconical flows, disc with hot spots, etc) that could be compared against the dataset presented here.
Since our fast GPU convolution implementation ($>1.28$ million convolutions across 5 days using gSTAR) and our data management infrastructure are already in place, this is a relatively straightforward task.
Further developments, like machine learning classification techniques, would provide additional tools to investigate the effect of second order source characteristics.

Our choice of 1.5 $R_{\rm Ein}$--long light curves, and 2000 light curves per source and map was justified in Sections \ref{sec:method} and \ref{sec:results} respectively.
The main restriction on the length and number of light curves is the storage space currently available on the host facility.
Whereas much of the microlensing analysis process (to date) has been compute or memory limited, we are now in a regime where we are I/O limited.
We are separately investigating the usefulness of data compression techniques (lossless and lossy) which may allow us to add additional light curves to the database \citep{Vohl2015}.

Planning of future and ongoing monitoring campaigns of specific systems could be facilitated by using our high resolution data and the \texttt{GIMLET} tool presented in Section \ref{sec:tools}.
Existing GERLUMPH maps for the $\kappa,\gamma$ values of the targeted systems can be selected from the GERLUMPH database and inspected.
By providing a value for $R_{\rm Ein}$ and the effective source velocity, low resolution light curves can be produced interactively by changing their length and orientation on the map.
The user can then intuitively decide the best observational strategy, e.g. in terms of the cadence of observations, based on the visual appearance of the light curves.

The \texttt{GIMLET} tool presented in this paper is a building block for a more advanced online modelling tool.
GPU acceleration in the browser provided by WebGL presents opportunities to model three--dimensional source profiles interactively in real time, e.g. to add components like an event horizon or jets, and modify their properties, like the opening angle or Schwarzschild radius.
WebGL could also be used to perform the convolutions between the modelled source profiles and the scaled down version of the map without any overhead, since all the computations would be performed in the browser in real time.
Finally, we intend to integrate the browser--based front--end more fully with a back end supercomputer, where user requests for high resolution full scale modelling would be sent for computation.
This working model may in fact not lie far from data access and analysis approaches that will be followed by the future all-sky survey facilities.

In conclusion, we have presented high quality microlensing simulations throughout the parameter space, preparing the theoretical ground for the upcoming all--sky survey era of quasar microlensing.
This dataset can be used as a benchmark for existing and future single object and parameter space studies, and can be further explored using machine learning and data mining techniques.
Our data and software is made openly available for further use by the community.
We complement our data with comprehensive and innovative online analysis tools.
We hope that this work will contribute to the advancement of data intensive quasar microlensing studies of the future.

\acknowledgements
This research was undertaken with the assistance of resources provided at the GPU--Supercomputer for Theoretical Astrophysics Research (gSTAR) through the Astronomy Supercomputer Time Allocation Committee, supported by the Australian Government.
gSTAR is funded by Swinburne University and the Australian Government's Education Investment Fund.
DC acknowledges receipt of a QEII fellowship from the Australian Research Council.

\bibliographystyle{apj}
\bibliography{gerlumph2}

\end{document}